# Size Effect of Diagonal Random Matrices

**A.A. Abul-Magd and A.Y. Abul-Magd**
*Faculty of Engineering Science, Sinai University, El-Arish, Egypt*

## Abstract

The statistical distribution of levels of an integrable system is claimed to be a Poisson distribution. In this paper, we numerically generate an ensemble of $N$ dimensional random diagonal matrices as a model for regular systems. We evaluate the corresponding nearest-neighbor spacing (NNS) distribution, which characterizes the short range correlation between levels. To characterize the long term correlations, we evaluate the level number variance. We show that, by increasing the size of matrices, the level spacing distribution evolves from the Gaussian shape that characterizes ensembles of 2×2 matrices tending to the Poissonian as $N \to \infty$. The transition occurs at $N \approx 20$. The number variance also shows a gradual transition towards the straight line behavior predicted by the Poisson statistics.

## 1. Introduction

Random matrix theory [1, 2] provides a framework for describing the statistical properties of spectra for quantum systems, whose classical counterpart is chaotic. It models the Hamiltonian of the system by an ensemble of $N$-dimensional random matrices, subject to some general symmetry constraints. For example, time-reversal-invariant quantum systems are represented by a Gaussian orthogonal ensemble (GOE) of random matrices when the system has rotational. A complete discussion of the level correlations for a GOE is a difficult task. Most of the interesting results are obtained for the limit of $N\to\infty$. Analytical results have long ago been obtained for the case of $N$=2 [3].It yields simple analytical expressions for the nearest-neighbor-spacing (NNS) $P(s)$, renormalized to make the mean spacing equal one. The spacing distribution for a two-dimensional GOE given by

$$p(s) = \frac{\pi}{2} s e^{-\frac{\pi}{4}s^2}, \tag{1.1}$$

is known as Wigner's surmise.The two-dimensional GOE obviously ignores the long range correlations within the spectra of chaotic systems. In spite of this limitation, the Wigner surmiseprovides a surprisingly accurate representation for NNS distributions of large matrices.

Berry and Tabor [4] conjectured that the fluctuations of quantum systems whose classical counterpart is completely integrable are the same as those of an uncorrelated sequence of levels. An infinitely large independent-level sequence can be regarded as a Poisson random process. The NNS distribution is given by

$$p(s) = \exp(-s). \tag{1.2}$$

An integrable system in quantum mechanics has, in principle, a known complete set of eigenvectors. The Hamiltonian matrix will naturally be diagonal in the basis that consists with this set. It is thus reasonable to model the integrable systems by an ensemble of diagonal random matrices. Interestingly, the NNS distribution derived from a 2×2 random matrix model is Gaussian and not Poissonian. This suggests that the limit of large $N$ is reached in integrable systemsmuch later that in the chaotic systems. The purpose of this paper is to estimate the minimal size of the random matrix ensemble that may be used to model large quantum systems.

In the present paper, weconsider both the short and long term correlations between levels characterized by the nearest-neighbor spacing (NNS) distribution $P(s)$ and the variance $\Sigma^2$, respectively, to discuss the eigenvalues statistics of $N$ dimensional diagonal random matrices. First, we show that the formula of NNS distribution of ensembles of the 2×2 random matrix ensemble has a Gaussian shape that characterize. An analogous derivation was given before by Chau [5] and Berry [6]. Thenwe numerically discuss the statisticspreviously mentioned above and compare the result with Poisson distribution and the distribution of Gaussian shape by 2×2 matrices.

## 2. Ensembles of Gaussian 2×2 Matrices

By consider a (2×2) real symmetric matrix

$$H = \begin{pmatrix} H_{11} & H_{12} \\ H_{21} & H_{22} \end{pmatrix}, \tag{2.1}$$

where $H_{11}$, $H_{12}$, $H_{21}$ and $H_{22}$ are real Gaussian random numbers with zero mean and variance $\sigma_{ij}^2$, and $H_{12}=H_{21}$.

The joint probability distribution for the matrix elements is

$$P(H) = \frac{1}{\sqrt{(2\pi)^2}\sqrt{2\sigma_{11}^2\sigma_{12}^2\sigma_{22}^2}} \exp\left[-\left(\frac{H_{11}^2}{2\sigma_{11}^2} + \frac{H_{12}^2}{2\sigma_{12}^2} + \frac{H_{22}^2}{2\sigma_{22}^2}\right)\right]. \tag{2.2}$$

In integrable systems, where the dynamical motion is integrable, the state functions are known in principle. They can be used as a basis for the matrix elements of the Hamiltonian. In this case, the Hamiltonian can be represented as a diagonal matrix. We shall therefore consider. the case when the diagonal elements have equal variances $\sigma_{11}^2 = \sigma_{22}^2 = \sigma^2$.In this case, Eq. (2.2) reads

$$P(E_1, E_2) = \frac{1}{4\pi\sigma} \exp\left( -\frac{E_1^2 + E_2^2}{2\sigma^2} \right). \tag{2.3}$$

Introducing the new variables $E = (E_1 + E_2)/2$and $s = E_1\text{-}E_2$, and imposing the condition of a unit mean spacing, the NNS distribution becomes

$$p(s) = \frac{2}{\pi} \exp\left( -\frac{s^2}{\pi} \right). \tag{2.4}$$

This distribution is not exactly the Poisson distribution $P(s)$ =exp $(-s)$.The 2×2 random-matrix model cannot describe the NNS distributions for large regular systems, which are known to be described by the Poisson distribution.

## 3. Numerical analysis

We numerically generate an ensemble of $N$-dimensional diagonal random matrices whose entries are pseudorandom values drawn from the standard normal distribution. In other words, we shall use a random number generator to generate values of the matrix elements so that they have a Gaussian probability density function.

For a physical system, the level density depends on the properties of the system under consideration. It varies from one system to another. One of the achievements of quantum chaology is that the fluctuation properties of energy spectra are universal when the spectra are "unfolded". The same is assumed for systems having regular classical dynamics. Unfolding consists in separating the secular variation from the oscillation terms. For this reason, unfolding is used to generate a spectrum whose mean level density is 1. The sequence of eigenvalues of each matrix of the ensemble generated according to the above procedure, $\{E_1, E_2 \ldots E_N\}$, after ordering does not have uniform average level density. To analyze the fluctuation properties, this spectrum has to be unfolded, i.e. specific mean level density must be removed from the data [7].

Every sequence is taken from the eigenvalues when unfolded, is transformed into a new sequence with unit mean level spacing. This is done by fitting a theoretical expression to the number *N* (*E*) of levels by use cubic polynomial or a Gaussian [8].

## I. Short range correlation between levels

The nearest neighbor spacing distribution $p(s)$ is the observable most commonly used to study the short–range fluctuations in the spectrum. This function is equal to the probability density that two neighboring levels $E_n$ and $E_{n+1}$have the spacing $s_n$. We calculate the NNS distributions of the eigenvalues for different *N* dimensional beginning from $N$=8 to arrive in the final to $N=10^4$ (~ ∞). The size of each ensemble is taken so that the total number of its eigenvalues is equal to $10^4$. For example, in one of the cases, we generate 1000 matrices of the size $N$=10.

Figures (1-a, b) shows the NNS distributions calculated for different values of *N* dimensional of diagonal random matrices. The figure shows a gradual transition of the shape of the NNS distribution from the Gaussian form (dashed line) to the Poissonian (solid line) as *N* increases. We observe a good agreementwith Poisson distribution at large *N*, Nevertheless for small dimensions we found statistically reliable to the Gaussian shape of 2×2. This is more clearly seen in Fig. (2), which reports the corresponding $\chi^2$ deviations, defined by

$$\chi^2 = \frac{4}{N}\sum_i \left( \frac{P_{\mathrm{P}}(s) - P_i(s)}{P_{\mathrm{P}}(s) + P_i(s)} \right)^2 , \qquad (3.1)$$

Where $P_{\mathrm{i}}(s)$ are the results of the numerical calculations, and the predictions of the Poisson distributions $P_{\mathrm{P}}(s)$ (black circle). The deviation from the Gaussian distributions are similarly defined and given by (white circle) in the figure.

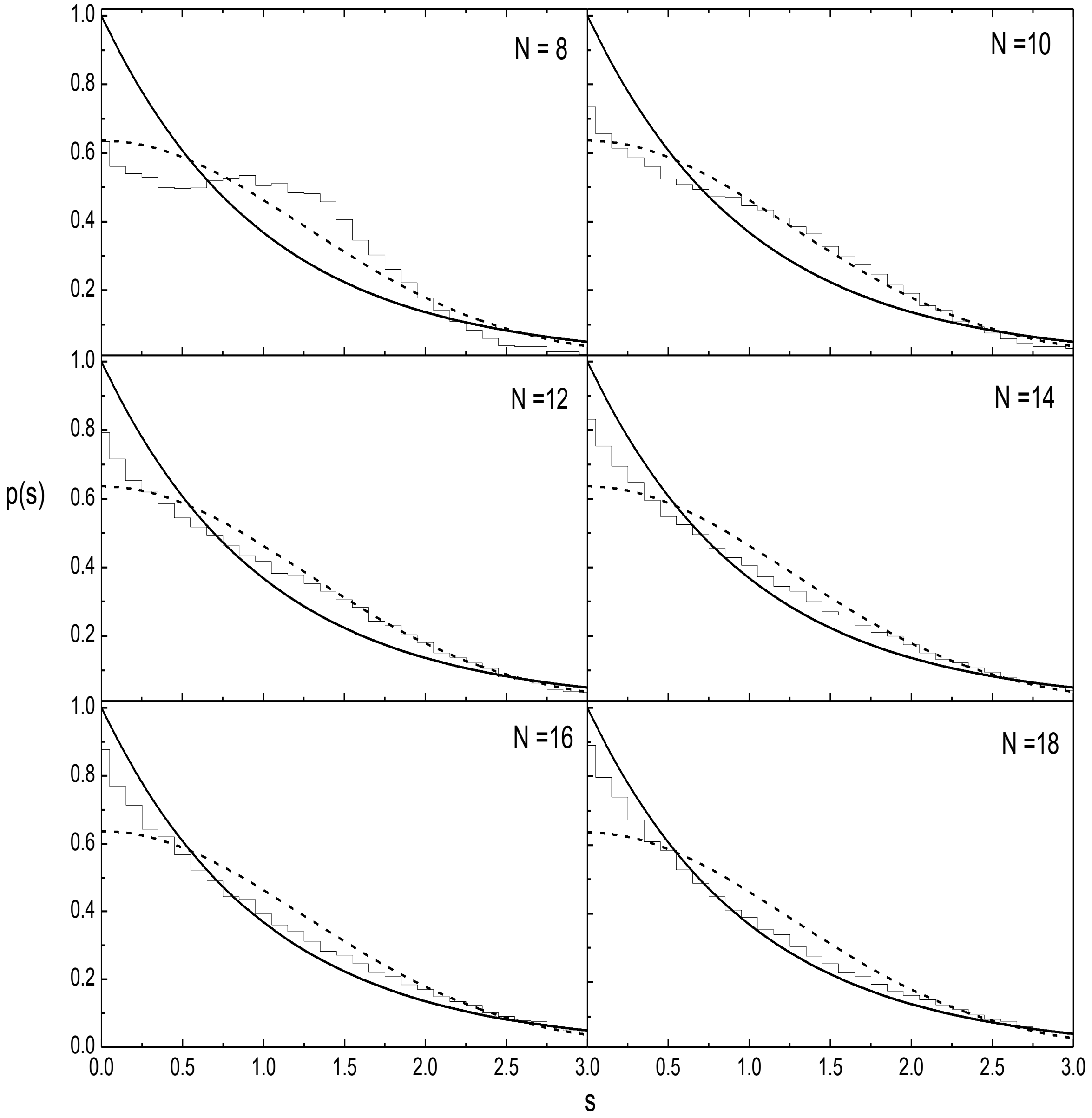

Fig. (1-a) the NNS distribution of Gaussian random diagonal matrices for $N$=8, 10, 12, 14, 16 and 18 as (step line), with the Gaussian shape of 2×2 (dashed line) and the Poisson distribution (solid line).

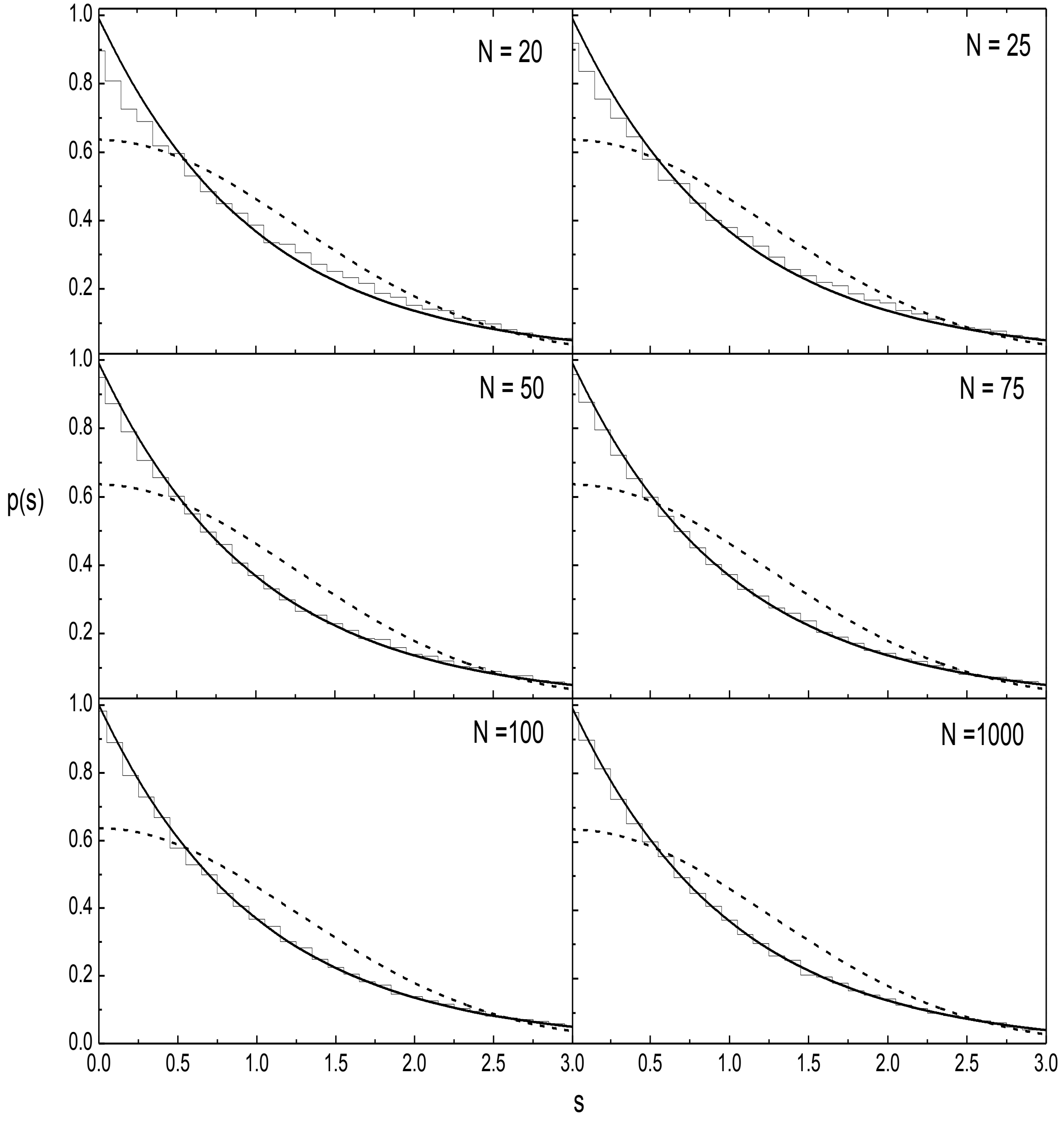

Fig. (1-b) the NNS distribution of Gaussian random diagonal matrices for *N*=20, 25, 50, 75,100 and1000as (step line), the Gaussian shape of 2×2 (dashed line) and the Poisson distribution (solid line).

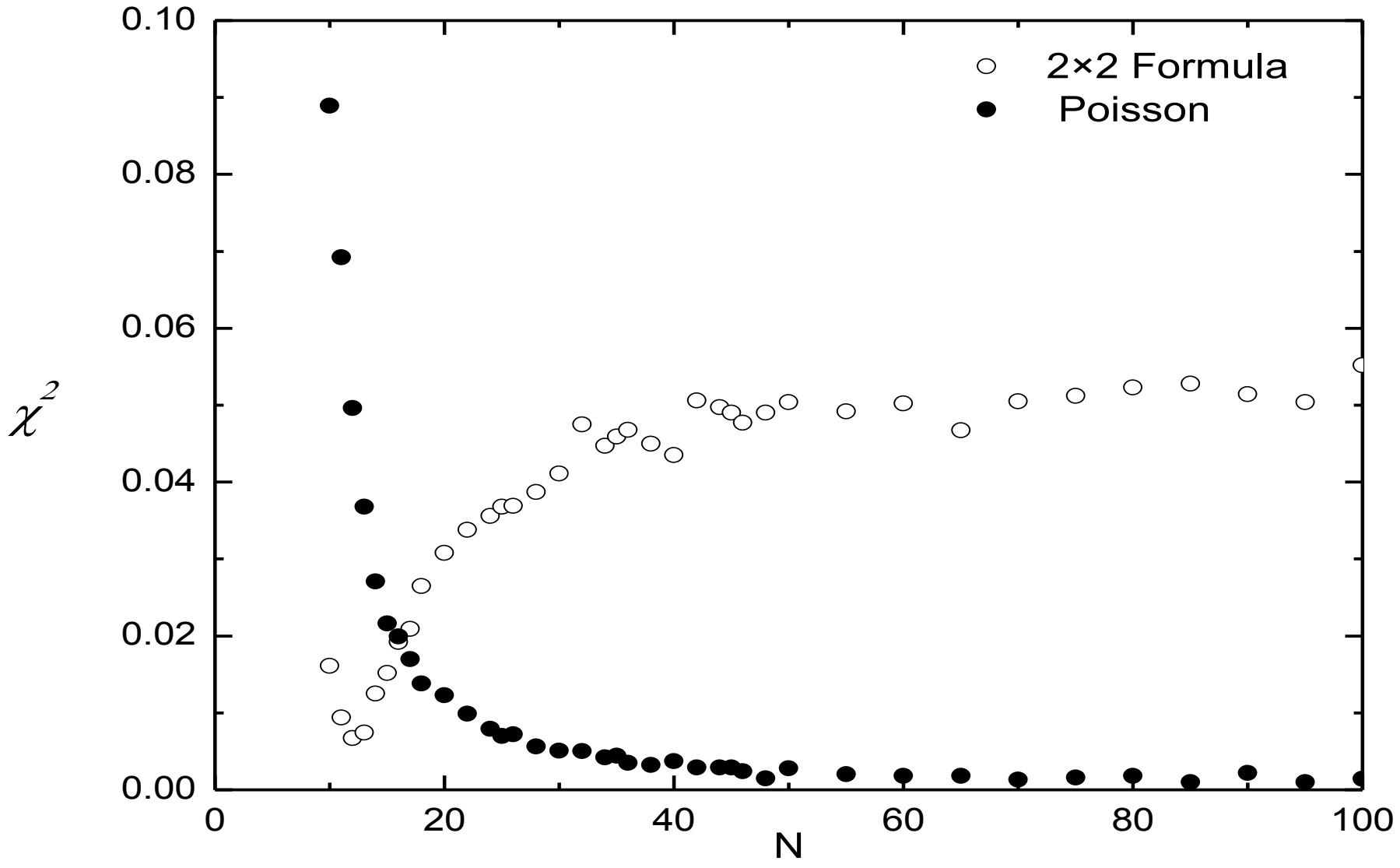

Fig. (2) $\chi^2$ for the *N* dimensions of Gaussian random diagonal matriceswith the Poisson distribution(black circle)and 2×2 formula definition(white circle).

## II. Long range correlations between levels

The $\Sigma^2$statistic characterizes the long ring-term correlation between levels, being the level-number variance of the spectrum. Specifically, for a given number $L$ of levels the variance [9] of the number $n(L,\varepsilon)=N(\varepsilon+L)-N(\varepsilon)$ of unfolded energy-levels in the interval $[\varepsilon, \varepsilon+L]$

$$\Sigma^2(L,\varepsilon) = \left\langle [N(L,\varepsilon)-L]^2 \right\rangle_{\varepsilon}, \qquad L>0. \tag{3.2}$$

Where $N(\varepsilon)$ is the integrated density of unfolded eigenvalues and $\langle\ \rangle_\varepsilon$ denotes an average over $\varepsilon_0$. Our purpose is to calculate the value of $\Sigma^2(L)$ numerically for the ensembles of the diagonal random matrices deferent sizes, which have been calculated in the previous subsection.

Fromthe numerical calculations mentioned above, weevaluate the level number variancefor the eigenvalues of the matrices of ensembles having different sizes and then compare the result with the variance of Poisson statistics.

The results so obtained for $\Sigma^2(L)$ have been plotted against $L$ in Fig. 3). The straight linecorresponds to behavior of $\Sigma^2(L)$ for a Poisson statistics of the level spacing as:

$$\Sigma^2_{\text{Poisson}}(L) = L \tag{3.3}$$

The dashed line and doted line in the figure showed $\Sigma^2(L)$ for GOE and GUE as:

$$\Sigma^2_{\mathrm{GOE}}(L)=\frac{1}{2\pi^2}\left[\ln(2\pi L)-\cos(2\pi L)-\mathrm{Ci}(2\pi L)+\gamma+1+\frac{1}{2}\mathrm{Si}^2(\pi L)\right.$$

$$\left.-\frac{1}{\pi}\mathrm{Si}(\pi L)+2L\left(1-\frac{2}{\pi}\right)\mathrm{Si}^2(2\pi L)\right], \qquad (3.4)$$

and

$$\Sigma^2_{\mathrm{GUE}}(L)=L+\frac{1}{2\pi^2}\left[\ln(2\pi L)-\cos(2\pi L)-\mathrm{Ci}(2\pi L)+\gamma+1-2\pi L\mathrm{Si}(2\pi L)\right], \qquad (3.5)$$

where Si($L$) and Ci($L$) are the sine- and cosine-integral functions, respectively, and $\gamma$ is Euler's constant [10].The Figure showed the agreement of the variance for lower size matrices with the variance formula of Gaussian ensemble, otherwise when increase the dimension of matrices the variance shows a gradual transition towards the straight line behavior predicted by the Poisson statistics.

This is more clearly seen in Fig. (4), which reports the corresponding $\chi^2$ deviations as (3.1) between the results of the numerical calculations of $\Sigma^2$ statistics of $N$ dimensions and the predictions of the Poisson distributions (3.3). The deviation from the Gaussian distributions are similarly defined and given by (3.4, 5) in the figure.

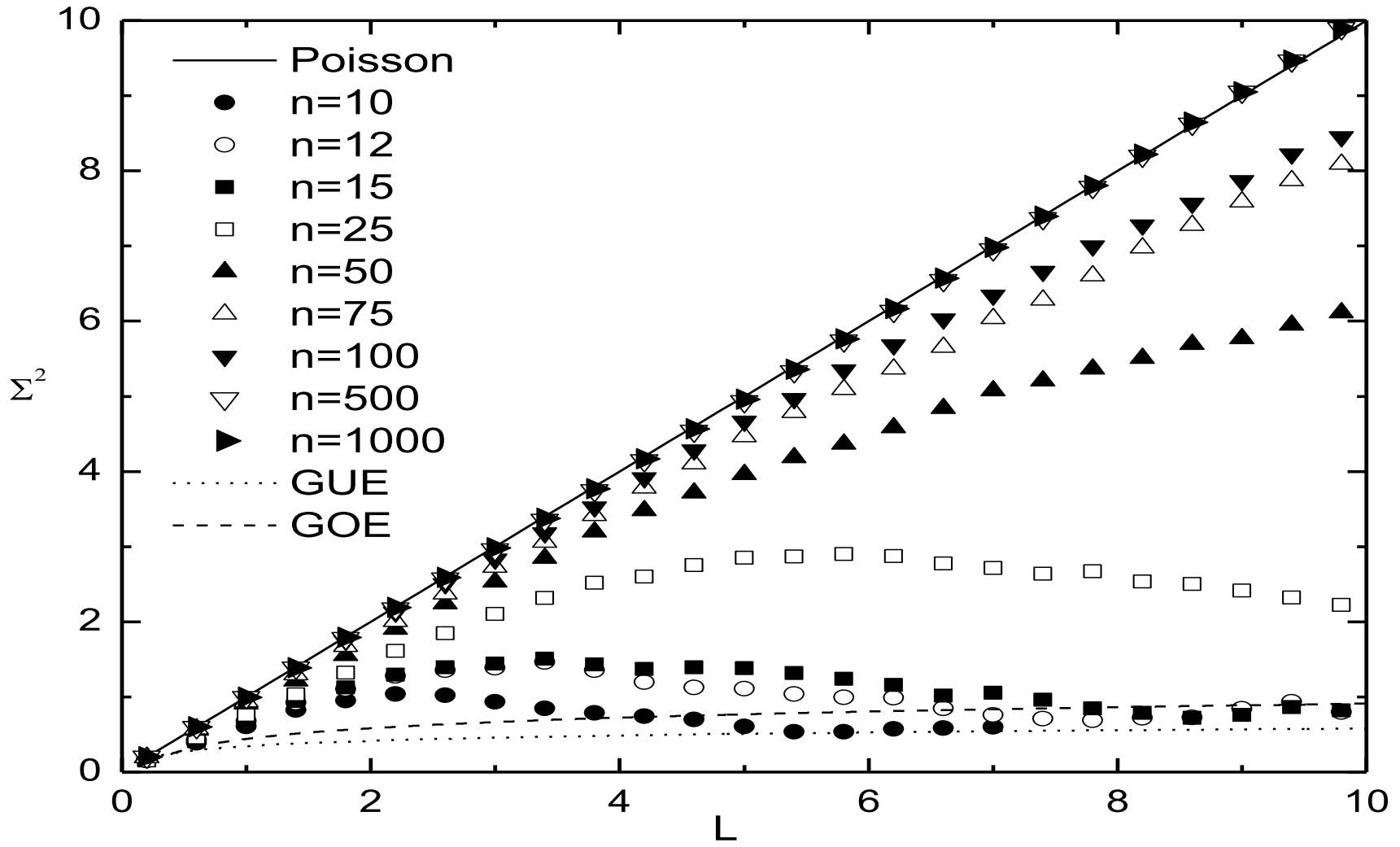

Fig. (3) $\Sigma^2(L)$ for $N$=10, 12, 15,25,50,75,100,500 and 1000, for the Poisson distribution (solid line) and for the Gaussian shape of GOE (dashed line) andGUE (dotted line)

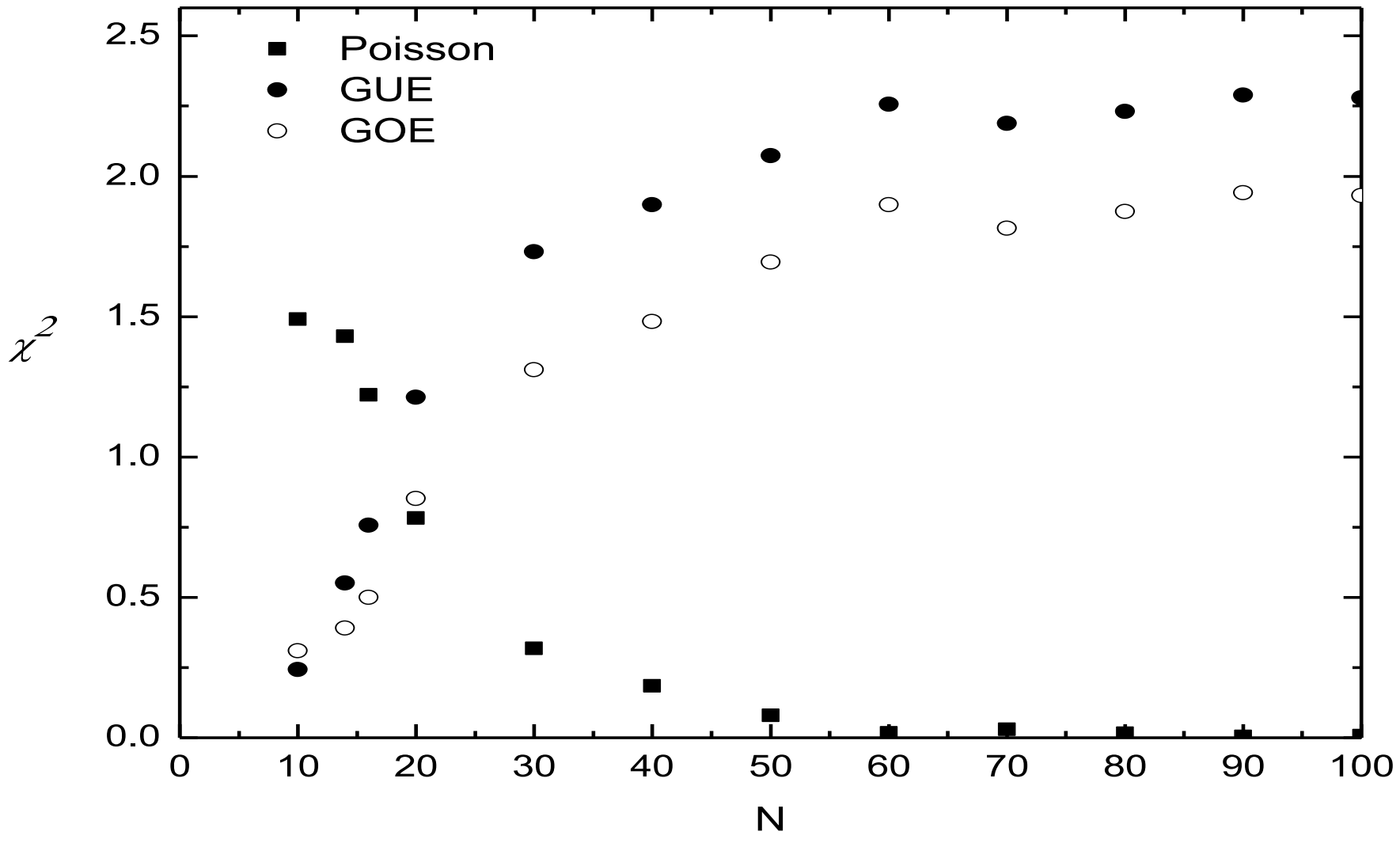

Fig. (4) $\chi^2$ for The number variance of the $N$ dimensions of Gaussian random diagonal matriceswith the Poisson distribution (black Rectangle) and GOE formula definition (white circle),the GUE formula definition of the number variance Represented by (black circle).

## 4. Discussion

Most of analytical discussion ofdevoted to mixedregular-chaotic systems using random matrix models are obtained for 2×2 matrices by severed authors (e.g. [7] and references therein).Winger [11] startsthis discussion with random matrices that model chaotic systems.This givesa NNS distributionthat almost coincides with the one obtained by numerical methods for matrices with $N$>> 1.In the case of integrable system random matrices the situation is different. The Gaussian behavior of the NNS distribution predicted for the $N$=2 case gradually modified to a Poisson distribution as $N$ increases. We here note that Drukker and Gross [12] have considered the expectation value $\langle \mathrm{Tr}\ \exp[\mathrm{M}] \rangle$, where M is an $N$×$N$ Hermitian matrix randomly chosen from a GUE with standard deviation $\sigma$. They evaluated the expectation by expressing the Vandermonde determinant in terms of the Hermite polynomials and then expressed the resulting expression as an expansion in powers of $1/N$. They showed that, while for any finite $N$ the quantity $\langle \mathrm{Tr}\ \exp[\mathrm{M}] \rangle$,behaves as ~exp(-$\sigma^2$), it behaves at large $N$ as ~exp(-$\sigma$).

This paper considers a Regular system, which is modeled by setting the off-diagonal matrix element of a GOE equal to zero, as previously done by several authors. This results in a Gaussian NNS distribution that does not fully agree with the Poisson distribution. This Gaussian behavior is valid for the $N$=2 case and may very well be modified to a Poisson distribution at large $N$.We show a gradual transition towards Poisson distribution by increasing $N$. The transition parameter occurs approximately at $N \approx 20$.

## References

[1] M.L. Mehta, Random Matrices 2nd ed., Academic Press, New York, 1991.

[2] T.Guhr, A. Müller-Groeling, H.A. Weidenmüller, Phys. Rep. 299, 189 (1998).

[3] C.E. Porter, Statistical Properties of Spectra: Fluctuations, Academic Press, New York, 1965.

[4] M.V. Berry, M. Tabor, Proc. R. Soc. Lond. A 356, 375 (1977).

[5] P. Chau Huu-Tai, N.A. Smirnova, P. Van Isacker, J. Phys. A: Math. Gen. 35, L199 (2002).

[6] M. V.Berry, P. Shukla, J. Phys. A: Math. Theor. 42, 485102 (2009)

[7] Bertuola A. C., de Carvalho J. X., Hussein M. S., Pato M. P. and Sargeant A. J., arXiv:nucl-th/0410027v2 (2005).

[8] A. Y. Abul-Magd and H. A. Weidenmüller, Phys. Lett. 162B, 223 (1985).

[9] R. Aurich, A. Bäcker, and F. Steiner: Int. J. Mod. Phys. B11, 805 (1997)

[10] Berry, M V, eds. S Albeverio, G Casati and D Merlini, Springer Lecture Notes in Physics, No.262, 47-53. 1986

[11] Wigner E.P., Can. Math. Congr. Proc., University of Toronto Press, Toronto, p. 174,(1957).

[12] N. Drukker, D.J. Gross, J. Math. Phys. 42, 2896 (2001).